# On Classification Issues within Ensemble-Based Simulation Tasks

Sergey V. Kovalchuk, Aleksey V. Krikunov, Konstantin V. Knyazkov,
Sergey S. Kosukhin, Alexander V. Boukhanovsky

*ITMO University, Saint Petersburg, Russia*
*kovalchuk@mail.ifmo.ru, alexey.v.krikunov@yandex.ru, constantinvk@gmail.com,*
*skosukhin@gmail.com, avb_mail@mail.ru*

**Abstract**
Contemporary tasks of complex system simulation are often related to the issue of uncertainty management. It comes from the lack of information or knowledge about the simulated system as well as from restrictions of the model set being used. One of the powerful tools for the uncertainty management is ensemble-based simulation, which uses variation in input or output data, model parameters, or available versions of models to improve the simulation performance. Furthermore, the system of models for complex system simulation (especially in case of hiring ensemble-based approach) can be considered as a complex system. As a result, the identification of the complex model's structure and parameters provide additional sources of uncertainty to be managed. Within the presented work, we are developing a conceptual and technological approach to manage the ensemble-based simulation taking into account changing states of both simulated system and system of models within the ensemble-based approach. The states of these systems are considered as a subject of classification with consequent inference of better strategies for ensemble evolution over the simulation time and ensemble aggregation. Here the ensemble evolution enables implementation of dynamic reactive solutions that can automatically conform to the changing states of both systems. The ensemble aggregation can be considered within a scope of averaging (regression way) or selection (classification way, which complement the classification mentioned earlier) approach. The technological basis for such approach includes ensemble-based simulation techniques using domain-specific software combined within a composite application; data science approaches for analysis of available datasets (simulation data, observations, situation assessment, etc.); and machine learning algorithms for classes identification, ensemble management and knowledge acquisition. Within the work, a set of case studies is addressed to examine the opportunities provided by the developed approach: metocean events' forecasting simulation, urban traffic environment, multi-agent crowd simulation, etc.

## 1 Introduction

One of the important issues within a context of complex system simulation is uncertainty management [1]. The uncertainty may come from different sources: lack of information about the simulated system, imperfect knowledge, imprecise data, restrictions of the model set being used. Ensemble-based simulation is often considered as a tool for management of uncertainty in various problem domains: hydrometeorology [2], life sciences [3], biology [4], etc. This approach is based on variation in input or output data, model parameters, or available versions of models to improve the simulation performance. Still the approach requires additional enhancement of the simulation process to manage the ensemble. Moreover, the consideration of the dynamically changing system may lead to the emergence of the evolutionary approaches implemented in the ensemble management procedures, where the ensemble is considered as a set of system's states, or variated data (parameters, input or output data), or even models. The work [5] presents the developing general purpose evolutionary approach to manage ensembles within various tasks of complex system simulation. One of the issues within the generalized ensemble-based simulation is the management of the ensemble within a cycle of diversity creation – uncertainty estimation – ensemble aggregation. In general, the ensemble aggregation can be performed in a two different ways: regression (combination of the ensemble elements) or classification (selection of a sole ensemble member as a result). The regression approach is the most popular one among the ensemble-based solution developers. Still its weakest point in many cases is that it may lose extreme values of original ensemble elements while in many cases these values are of especial importance (e.g. forecasting of extreme events – floods, hurricanes, earthquakes, etc.). In that case, the classification approach can be employed to overcome this issue. Within the presented work, we try to organize the existing practices of ensemble-based classification on the basis of generalized ensemble-based simulation framework [5] and extend the framework with a conceptual basis for classification-based ensemble management.

## 2 Related Works

Today the ensemble-based techniques are widely used within diverse areas of science. An ensemble can be considered in a various ways that differ by the procedures of ensemble building, diversity control, management during the simulation process and aggregation of the ensemble.

One of the popular ways of ensemble building is *multi-model* approach where different models are combined to provide alternative or competitive solutions for the task. There are several works consider ensemble of different evolutionary models with possible exchange between populations within ensemble: e.g. work [6] consider the ensemble of various discrete differential evolution (DDE) algorithms using the generalized traveling salesman problem (GTSP) as a benchmark; work [7] presents an ensemble of smart bee algorithms for optimization of large scale power systems; work [8] describe generalized approach for ensemble of constraint optimization algorithms with different approaches for constraint handling. Still the models within the ensemble can be of different nature: e.g. work [9] uses a set of artificial neural networks (ANN) built using various approaches for weather prediction. The multi-model approach is widely used for hydro-meteorological and climate simulation, where the external models can be often used [2], [10]. Moreover in many cases multiple data sources can be considered as a specific kind of multi-model ensemble: work [11] uses different sources (radars, meteorological stations, satellites, etc.) of precipitation data for hydrological processes simulation in Rijnland area (Netherlands). It is noteworthy that many of the mentioned cases use either one approach or one algorithm or even one model for building the ensemble. Still the ensembles can be considered as multi-model ensembles as the structure of member models is the subject to change.

*Ensemble learning* solutions usually can be treated as a specific type of multi-model ensembles as they consist of different classifiers or predicting functions. There are many techniques for ensemble management applied in machine learning (ML) tasks (e.g. bagging, boosting) used to manage and improve ensemble of ML task solvers. Usually the ML tasks consider classification [12], [13] or regression [14], [15] problem that can be reflected on the ensemble aggregation approach: selection of a single instance for classification and combination of the existing instances for regression.

A different approach for ensemble based simulation is based on model parameter variation, which can serve for different purposes. The parameter variation can be used for stochastic simulation with *stochastic parameters*. For example, work [16] uses ensemble simulation of this kind for global prediction of $CO_2$ emissions. A stochastic-dynamic parameterization approach [17] also can be considered within this scope being devoted to estimating sub-grid scale features of the modeled system. A stochastic ensemble is used to simulate global infection spread taking into account sociodemographic and population mobility in [18]. Variations of the stochastic parameter approach can include perturbed parameter ensembles and stochastic-dynamic parameterization [19]. Additionally this approach becomes especially important in case of uncertainty management issue [10], [20], [21], [22], dynamic adaptation to changing external condition [23], [24], or ensemble data assimilation (e.g. ensemble Kalman filter (EnKF) applications [19], [25]).

Another way of parameter variation is initiated by the goal of *parameter space coverage*. Work [15] solves the task of real-time exploration of discrete event simulation results by the use of preliminary ensemble-based simulation with parameter space coverage using Latin Hypercube Sampling (LHS). The simulation inputs and outputs are analyzed by machine learning algorithm to provide the real-time results in interactive "what-if" manner. Parameter space division with predefined rules and linear models for each sub-region is used in [26] to provide predictive control of a hypersonic vehicle. Work [27] uses a set of possible combinations of model parameters covering the predefined parameter space region to give a probabilistic estimation of the wildfire growth on each time step during evolutionary optimization of the parameters. An ensemble of all possible Boolean Networks (BN) except for tautology or contradiction transition function is considered in [28]. A controllable ensemble of trajectories within parameter space is considered in [29] within Weighted Ensemble (WE) simulation.

Finally the *explicit parameter variation* with automatic or manual generation of parameter variation set can be used for different tasks (usually for parallel simulation of various scenarios). For example various configurations of parameters is used in [30] for ensemble simulation approach of vascular blood flow. Work [16] considers a set of scenarios for global $CO_2$ emissions prediction as a high-level ensemble. Automatic ensemble generation is hired in [31] to generate multiple particles set for simulation of pelagic organisms' migration in ocean environment taking into account multiple aspects of behavior and influencing factors.

Although there are many works within the ensemble-based simulation area including matured set of works on ensemble management in weather and climate simulation as well as ensemble learning techniques, the proposed multitude of approaches, principles and algorithms are mostly isolated within the tasks being solved, or in particular areas with a restricted class of models. The generalized approach for ensemble-based simulation that can cover various areas, levels, and principles of ensemble management is still absent. Within our current work we are trying to develop such generalized approach. Starting from the conceptual framework for ensemble-based simulation covered in [5] (briefly covered in Section 3.1) we are focused on the classification issues within ensemble-based simulation in the presented paper.

# 3 Formal backgrounds

This section presents an attempt to construct formalized conceptual basis for ensemble-based simulation and to consider the classification issues emerged within the scope of this process.

## 3.1 General purpose ensemble-based simulation

According to the basic conceptual claims proposed in [5] to develop a general-purpose approach for ensemble-based complex systems simulation we need to identify the basic operations that are involved in the ensemble management. To do this, we considered a three-layer conceptual framework (see Fig. 1), where layers are related to the investigated system $S$, data which describes the system $D$ and a model to simulate the system's behavior $M$.

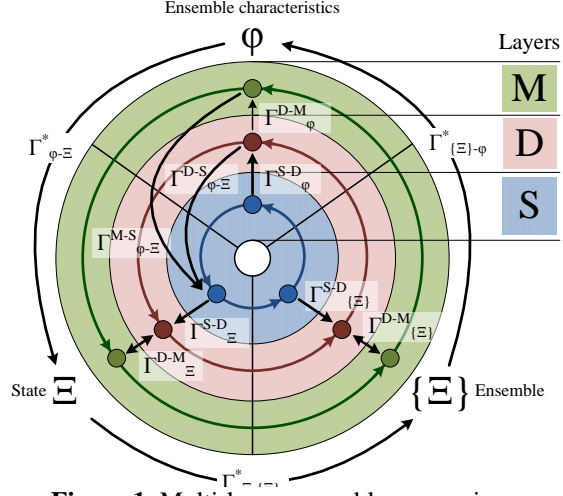

**Figure 1.** Multi-layer ensemble processing

Each layer includes the main artifacts that are involved into ensemble processing: (a) description (parameters) $\Xi$ of a single element related to the corresponding layer (state of the system $S$ dataset $D$ or model $M$); (b) an ensemble $\{\Xi\}$, considered as a set of elements, described earlier; (c) characteristics $\varphi$ of the ensemble to assess the evaluated ensemble and make conclusions on the ensemble analysis (one of the most important class of the ensemble characteristics is ensemble diversity characteristics).

To process these artifacts we can define a set of operators to represent moving from one artifact to another. Each operator is defined by two indices, denoting layer(s) and artifact(s): affected by the operator: $\Gamma^*_{\Xi-\{\Xi\}}$, $\Gamma^*_{\{\Xi\}-\varphi}$, $\Gamma^*_{\varphi-\Xi}$. These operators form a cycle which is often considered as a basic ensemble analysis procedure within a single layer: ensemble diversity creation, uncertainty analysis, and ensemble aggregation. To enhance the basic cycle of these three operators within the proposed multi-layer conceptual framework, we define additional operators required for the sake of consistency while different ensemble-based solutions are considered. These operators define the relationship between layers and influence of the higher layers onto the lower ones. Additionally the ensemble set depicted by the presented layers is evaluated over the time taking into account observation of the system to evolve the next generation of the ensemble. The basic framework is described with more details in [5].

### 3.2 Classification within ensemble-based simulation

The procedure of classification can be applied in various parts of the developed approach. First of all the classes set $C_t = C_t^{IMP} \times C^{EXP}$ where $C^{EXP}$ is a set of classes defined explicitly by the domain experts, $C_t^{IMP}$ is a set of classes implicitly identified after the available data analysis. The last set can be updated in an automatic way over the time by the operator $C_{t+1}^{IMP} = \mathfrak{K}_t(\Omega_t) C_t^{IMP}$. Then the classification operator $K_t: * \to C_t$ can be defined for class selection. Fig. 2 depicts the possible influence of classification result onto the operators within the ensemble evolution procedures: evolutionary operators that transform ensemble and classes set as well as layer operators within a single time step of evolution and classification operator on the next step.

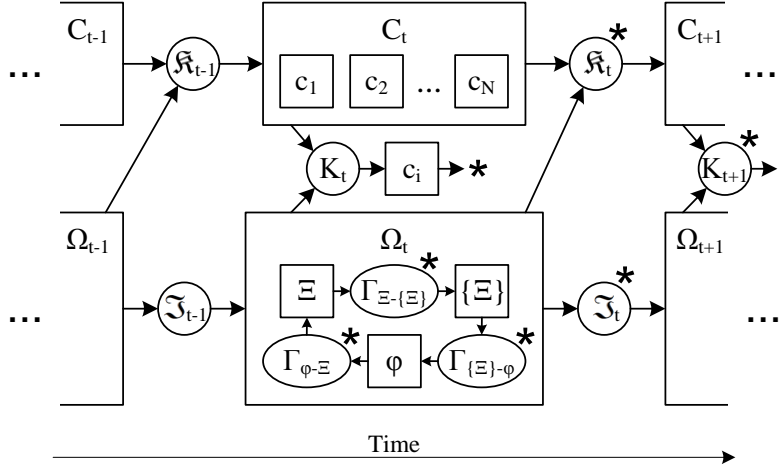

\* - possible influence of class selection
**Figure 2.** Classification and ensemble evolution

Within the developed approach, the operators $\mathfrak{K}$ and K can be applied in different ways. Here the simplest ways are mentioned:

1) *Basic classification.* Each layer's basic artifact (system, data, and model) can be used as an object for classification. Here $C^{EXP}$ can be considered, as a predefined set of classes, while $C_t^{IMP}$ usually is a result of clustering analysis. The classification can be applied to a) each element of the ensemble within the layer; b) result of ensemble aggregation. In any case, the result of classification can be used as a meta-data to enhance the artifact's processing by layer operators. The simplest way to do so is selecting the operators from a predefined sets or modification of the operators by passing the class as an operators' argument. First of all, it could be done for uncertainty estimation and aggregation operators.

2) *Classification of the ensemble state.* Additionally on each layer, the ensemble of artifacts and results of uncertainty estimation can be used to identify current class at certain time step. As well as in the previous case the identified class can be used to enhance operators applied to the ensemble.

3) *Classification-based aggregation.* The procedure of classification-based selection of ensemble's element can be used as an aggregation operator. Here the set of classes corresponds to the ensemble elements' set, and the estimation of ensemble's uncertainty allows selecting one of the elements. E.g. in case of an ensemble of models it can be supposed that $C_t = \{\Xi\}$, the operators $\Gamma_{\vartheta-\Xi}\Gamma_{\{\Xi\}-\vartheta}$ identifies the element selected on the current time step.

## 4 Unified procedure to build an ensemble

This section covers the basic procedures involved in the processes of ensemble building and controlling of its evolution with the use of available conceptual, algorithmic and technological toolbox.

### 4.1 Combination and selection of ensemble members

Mentioned earlier classification-based aggregation is one of the most useful strategies that can be implemented in various ways by application of different strategies of ensemble building and selection of the result member. In this section, a generalized approach to manage combinatorial ensemble by selection of best member is considered. The task of simulation-based forecasting is used as an example that can be mapped onto a wide range of problems in different problem domains.

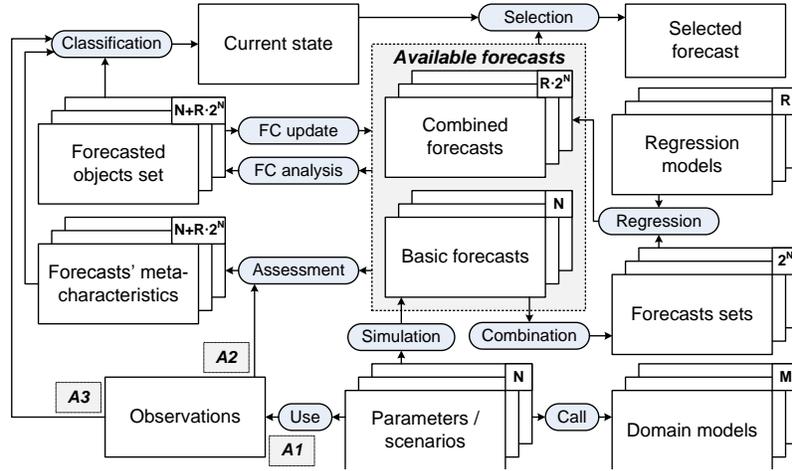

**Figure 3.** Combination and selection of ensemble members for simulation-based forecasting

The described approach includes several procedures that could be modified depending on the particular task's requirements. Fig. 3 shows the generalized procedures of ensemble management within a task of simulation-based forecasting and objects affected by these procedures (the number in upper right corner depicts the maximum available cardinality of an ensemble set).

1. *Building basic ensemble.* Usually, the simulation-based approach considers the ensemble elements as results of a simulation using a set of models or a single model with a set of parameters. In general approach, we can define a set of parameters as an initiative ensemble considering link to the model as one of the parameters. The elements of this ensemble can be either variated parameters applied to a single model ($M = 1$), or equivalent parameters applied to each of the models ($M = N$), or any combination of the models with a variation of parameters. Additionally some of the parameters may include the current or/and historical observations to be passed as assimilation parameters in case of the model supports the assimilation. The basic ensemble is built as a result of the simulation according to each of the parameters set in the initiative ensemble.
2. *Combination of the basic ensemble members.* Using the $N$ results of the simulation procedure, it is possible to build $2^N$ sets (including an empty set and a set containing all $N$ results). Each of the sets can be used to build an aggregated result (as a regression of the elements in a set). In case of an empty set, the result will be "agnostic" function (e.g. averaged historical value taken as a forecast). In case of sets with cardinality equal to 1, the result, for example, could be a statistical improvement of the forecasted data. As a result of this combination using $R$ forms of regression up to $R \cdot 2^N$ forecasts can be build. Still the easiest way of ensemble based simulation is the usage of a single (e.g. linear) combinatorial forecast built using a set of all $N$ results of the simulation.
3. *Forecast analysis and update.* Analysis of available forecasts (from $N$ up to $N + R \cdot 2^N$) has two goals: first of all, it assesses the available forecasts using observation data, predefined rules or interrelationship between the forecasts; secondly, analysis of the forecasts can identify high-level domain-specific objects within the forecasts (e.g. in case of sea level forecasting the floods can be considered as such high-level objects). Both these results (assessment results and identified high-level objects) can be used to update available forecasts directly (by modification of its values) or indirectly (by generation of combinatorial forecasts). Additionally the last variant enables cyclic improvement of the available forecasts set.
4. *Selection of the forecasts.* The classification-based aggregation is used to select the best one from available forecasts. The selection procedure can use all the information collected after the previous procedures: assessment results, identified high-level objects, observations, etc. The result of this procedure is a selection of a single forecast that can be either from the basic ensemble obtained by the simulation or from the combinatorial ensembles set.

One of the important procedures performed during the simulation is data assimilation, which considers the simulation improvement by modification of model parameters of input/output data. Within the proposed approach there are three major ways to apply the data assimilation procedure (mentioned in the Fig.3).

*A1) Direct assimilation* within the simulation with models that support this procedure. E.g., models can use currently observed data to modify initial conditions or update the forecast after the simulation.

*A2) Assimilation within forecasts assessment* can be performed to estimate the performance skills of the forecast by comparison the forecast to the available data (often at least the data for starting time point of the forecast is available) or by predicting the ensemble skill using observed data as one of the predictors.

*A3) Assimilation within selection procedure* can be applied to modify the selection procedure depending on the available observation data. E.g., different selection procedures can be applied for predefined ranges of currently observed values.

## 4.2 Tools for class identification and selection

The mentioned procedures can be implemented using various approaches. This section summarizes several popular approaches used on different layers of ensemble-based simulation (see Table 1). Usually, the tools on a lower level can be applied on a higher one as well (e.g. tools from system layer can be applied on data layer as data describe the simulated system, tools from data layer can be applied on the model layer as models could be considered as sources of data). Tools on each layer can be divided into two groups: explicit, which involve domain-specific knowledge to perform, and implicit, which try to infer the knowledge from existing data. This division is not strict since many of these tools involve both of the mentioned ways or at least can be considered from different points of view.

| **Layer** | **Explicit** | **Implicit** |
|---|---|---|
| System | - High-level objects identification<br>- Rules on observations | - Clustering of system state<br>- Automatic shape analysis, pattern recognition<br>- State switching control |
| Data | - Data sources meta-data analysis<br>- Rules on data sources<br>- Comparing to observations | - Data skill (quality) prediction<br>- Analysis of data set interrelationship |
| Model | - Predefined regression forms<br>- Rules on ensembles | - Symbolic regression<br>- Artificial neural network (ANN)<br>- Statistical analysis (e.g. principal component analysis, PCA)<br>- Ensemble skill (quality) prediction<br>- Analysis of ensemble set interrelationship |

**Table 1.** Classification and selection toolbox

On the system layer, the explicit knowledge can describe the structure of high-level objects that can be identified within simulated data, and rules for switching selection procedure depending on observation values. The implicit way here includes various machine learning techniques to identify clusters, patterns, and states using available data. Additionally the control of state transitions can be applied here (e.g. keeping the state while the switch criterion didn't reach defined level).

Within the data layer multiple datasets that provide information about the simulated system are considered. Thus, the explicit analysis includes analysis of the meta-data for this sources (e.g. trusting level, confidentiality, etc.), comparing this data to available observations and extending the rule set with data-based rules. The implicit analysis involve prediction of data quality (which is especially important for forecast data) and estimation of various interrelationships between datasets (correlation, distances, inconsistency, etc.). This information can be used during class set building as well as during selection of available data set.

The model layer presents an analysis of the models that are used for simulation. The explicit knowledge beside model-based rules can include a form of regression defined by an expert, which can be based on the knowledge about models nature. Implicit toolbox here can be extended with several techniques that can control the ensemble structure by an analysis of models' behavior: symbolic regression, artificial neural networks or statistical inference (e.g. principal component analysis). Also, implicit classification procedures may include quality prediction and interrelationships analysis for models and ensembles of models.

## 4.3 Quality assessment

The analysis of quality of the ensemble members (and in turn, the whole ensemble) is an issue of significant importance for ensemble-based simulation. Usually, the quality assessment is based on an estimation of the distance between simulation results and corresponding observations (see A2 assimilation procedure in Section 4.1). Thus, the important questions arise: a) how to measure the distance within the system's state space; b) how to quantify the quality of the ensemble according to the selected metric?

The *distance measure* can vary depending on the particular problem: it can be general purpose metrics like Euclidean distance; probabilistic metrics like Mahalanobis distance, Chernoff distance etc. [32]; metrics based on information theory like metric based on Kolmogorov complexity [33]; specific distances for time series like Dynamic Time Warping (DTW), Longest Common Subsequence (LCSS) etc. [34]; or even exotic distances invented, for example, for string comparison like Levenshtein distance, Jaro distance etc. [35]. Within the scope of ensemble-based simulation the metric can be applied to measure the distance between observation data and simulated data as well as distance between different simulation results (either within the ensemble or between the ensembles). While the first case is mostly used to measure quality of the ensemble, the second one is involved in the diversity quantification, uncertainty management, sensitivity analysis, etc.

The *quality measure* is defined to compare different ensembles, to control the ensemble evolution and dynamic behavior and to optimize the ensemble management procedures. The functional quantification of ensemble quality can also vary from standard and well known approaches to techniques developed for the particular tasks. Some of the examples are: Mean Average Error (MAE), Root Mean Square Error (RMSE), Ignorance Score, Brier Score, Ranked Probability Score, etc. The developed assessment procedures can be based on probability metrics [32], statistical procedures [36], information theory [37], signal processing [38], etc.

Within the ensemble-based simulation the mentioned measures are used a) internally to manage the ensemble; b) externally to analyze the overall quality of the ensemble. It can be either the same metrics or different according to domain knowledge. Moreover, the metrics can be combined which leads to a multi-objective optimization problem. In any case, the selection of the measures has a significant influence on the vitality of the particular ensemble during the evolution as well as the implemented ensemble-based solution. Nevertheless, there is no definitive solution for measure selection as it depend significantly on the problem domain, particular task being solved, and models being used. Thus, the measure selection support is considered as one of open questions planned for future work within the presented conceptual approach.

## 5 Case study: flood ensemble forecasting

### 5.1 Problem statement

To demonstrate the mentioned ensemble-based techniques several applications within the use case of sea level forecasting are presented in this section. The task of Baltic Sea level forecasting has especial importance for protection of Saint Petersburg from floods. Saint Petersburg suffers from floods during all the history of the city: from the foundation in 1703 about 300 of floods were detected. Today the boundary of flood detection in the city is defined at the water level of 160 cm according to the gauge in the mouth of Neva River near the National Mineral Resources University (the highest registered flood in the city's history was detected in 1824 and was 421 cm of water level in the city). The floods in Saint Petersburg area (eastern part of the Gulf of Finland) is caused mainly by storm surges, which in turn are caused by cyclones traveling over the Baltic Sea [39].

To protect the city from floods, the Saint Petersburg Flood Prevention Facility Complex was put into operation in 2011. The complex has capability to prevent floods up to 5 meters high. Still the flood prevention requires the closing of 8 gates that are normally opened for ships and water passage. The development of the plan for gates operations is based on water level forecasts that are built using a complex set of models and can be characterized by uncertainty of different kind needed to be managed [40]. Moreover, the plan development procedure should take into account the technical characteristics of the gates, the flow of the Neva River, which rises water level in case the gates are closed, requirements of different stakeholders (sea port, emergency and rescue services, etc.), interaction with decision makers etc. [41]. Thus, the improvement of forecast quality is a significant task a) to predict floods as early as possible; b) to perform simulation-based assessment and optimization of developed plans and final decision-making support.

Within the research, we use a set of 13 alternative models. The basic set of 12 models was built using two alternative software packages (BSM and BALT-P) originally developed for Baltic Sea water level simulation. The software packages have various modes of use: different scales, alternative sub-models, optional internal data assimilation, etc. This enables alternative models building using a single software package. Additionally, both software packages use meteorological forecasts as input data for simulation. Several external sources of meteorological data was used namely GFS, HIRLAM, FORCE, COSMO. Within following descriptions, the forecast sources are encoded using the following pattern <model name>-<options>-<meteorological data source> (for example BALTP-90M-GFS means that the forecast was obtained using BALT-P model on the grid with 90-meter step and GFS data source as meteorological input). Additionally one external model (HIROMB) was used to extend the set. The result set of the model can be used to get a set of 13 water level forecasts for different length (from 60 hours to 192 hours) depending on the meteorological source with common inter-forecast time of 6 hours. The presented experimental study uses the presented model set to demonstrate the issues mentioned earlier.

### 5.2 Classification within ensemble

The basic approach of the ensemble based simulation is usage of a combination of original forecasts (the simplest form of combination is weighted sum of forecasts. The simplest way to do this is to combine the forecasts using a linear regression or more advanced techniques like primary component analysis (PCA) (see some examples in [5]). To analyze the influence of various classification-based procedures, as a reference model we used a linear combination of forecasts, where weighting coefficients and intercept coefficient are optimized using historical observations with the least squares method. This section covers different classification-based procedures aimed towards ensemble management improvement using the mentioned basic ensemble as a starting point.

*Basic selection.* In some cases, the selection procedure can alter from the basic ensemble to one of the source forecasts. The switching can be done after comparison with available observations using predefined heuristic rules. The easiest way to build such rules is to compare of distances between forecasts, ensemble, and observation at the beginning of the forecast (where the observations are often available). Switching can be performed if the ensemble stays much far from the observations, comparing to all source forecasts, or if the range of the forecasted value at the beginning covers observations and don't cover the ensemble, etc. E.g., in the case shown in fig. 4, switching from the basic ensemble to forecast from source BSM-BS-GFS can decrease mean average error (MAE) of this particular forecast by 9.2% (from 5.42 cm to 4.92 cm).

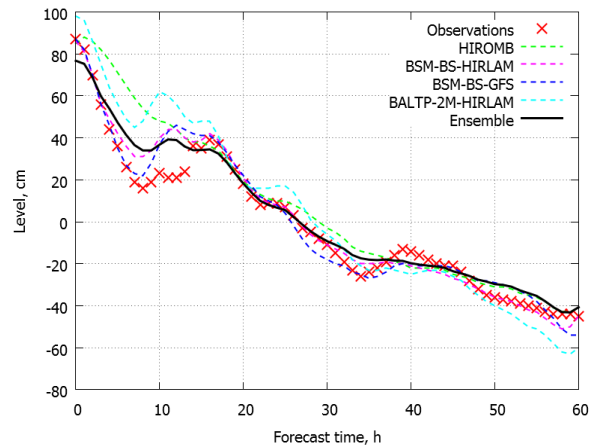

**Figure 4.** Selection of initial forecasts instead of ensemble forecast

*Selection of ensemble members.* Selection from a set of $2^N$ ensembles can improve the forecasting because a) ensembles with fewer members can show better performance (especially in cases of multicollinearity); b) the quality of forecast from data sources can vary in time. Thus, the changing of ensemble members set can improve the quality of the forecast. One of the ways of the selection procedure implementation is the estimation of the ensemble forecasts' quality and selection of the best one according to the estimation. As an example, fig.5 (left) shows a comparison of error (calculated as DTW distance to observations) for ensemble, containing all source datasets, to ensemble, selected with the proposed approach, after the estimation of error using linear regression on two observation history point, beginning from forecast starting time. The average improve is about 8.9%. An example showing a selection of better forecast is shown in fig. 5 (right). Still there are points where the implemented algorithm selects the ensemble that increases forecast error.

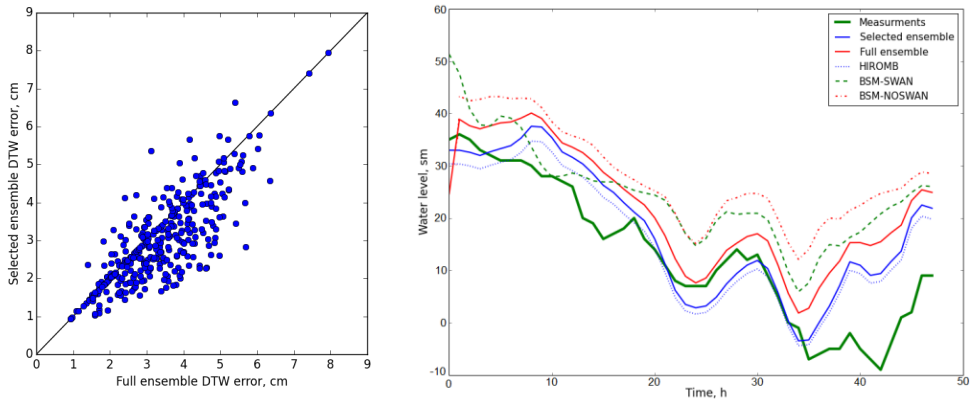

**Figure 5.** Selection of ensemble members: average improve (left); example of a forecast (right)

*Conservative selection.* In some cases, the selection of the forecast can be improved if the switch of the class is performed only if the new selection provides a solution with a quality that exceeds the current solution by the defined threshold or more. This enables more stable system state with less switching. Fig. 6 shows that conservative solutions outperform non-conservative with a threshold in a range [0.04; 0.96].

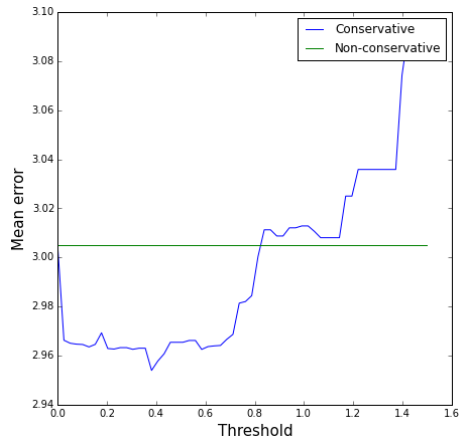

**Figure 6.** Conservative selection with predefined threshold

*Automatic function building.* One of the popular ways to identify the functions that can be applied as selectors or ensemble combination forms is symbolic regression [42]. E.g., in the proposed solution the form of the regression for two forecast sources obtained by the Pareto genetic programming are presented in fig. 4h. Here data sources are denoted by letters "h", "b" while C0…CN are free coefficients. The forms ## 0, 4, 5, 9 presents $2^N = 4$ linear ensembles, while the best scores are obtained for forms ## 6, 8, 10, 11, 12, which introduce additional operations on the ensembles. Nevertheless these forms need to be explained further from the point of view of domain experts (while, e.g. linear combination can be explained naturally). After the training, the simple Markov chain prediction of the best from these forms enables decreasing of mean average error by 9.4% (from 8.25 cm for linear regression of all elements to 7.47 cm).

| | Individual | #6 | (C0 ×h ×b-C1) |
|---|---|---|---|
| #0 | C0 | #7 | (C0 ×b-C1 ×h) |
| #1 | b | #8 | (C0-C1 ×max(b,h)) |
| #2 | h | #9 | ((C0 ×b-C1 ×h)-C2) |
| #3 | max(b,h) | #10 | (C0 ×max(C1 ×b,C2 ×h)-C3) |
| #4 | (C0 ×b-C1) | #11 | ((max(h,b)-min(C0 ×max(b,h),C1))-C2) |
| #5 | (C0 ×h-C1) | #12 | ((C0 × $\frac{(C1 \times h - C2)}{(b-100.000)}$ -C3)-max(C4 ×b,C5 ×h)) |

**Figure 7.** Symbolic regression for ensemble aggregation function

## 5.3 Domain-specific classification

The process of knowledge-based class identification can be performed either using explicit domain-specific knowledge or by identification of implicit knowledge during data analysis. Here we present two attempts to use the explicit (level peak description) an implicit (automatic identification of forecast anomalies) knowledge to improve ensemble-based simulation.

*Flooding peak parameterization.* The most important capability of the forecast is the prediction of floods that are characterized by level rise over 160 cm. To enhance the floods forecasting, the high-level object was introduced to parametrize peak of the level higher than the predefined level (which can be lower than flood level). We introduce four parameters to characterize peak within forecast (see fig. 8): $H$ – maximum level within the peak; $T$ – time from forecast start to the maximum level; $W$ – width of the peak (time from crossing the threshold when level goes up $T_L$ to crossing it when level goes down $T_R$); $D$ – ratio $(T - T_L)/W$. The peaks can be identified in each of the forecasts as well as in observation data.

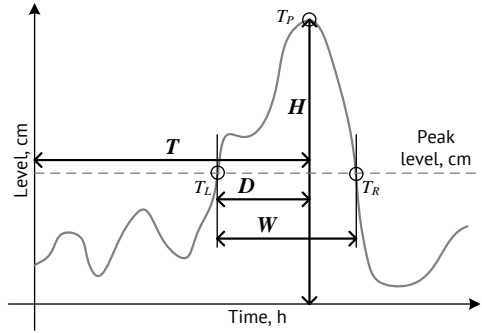

**Figure 8.** Sea level peak parameterization

Then the forecasts ensemble transformation is applied (see Fig. 9 (left) for example forecast with threshold level 120 cm; (1) – observations; (2) – original forecasts from three data sources; (3) – ensemble built using a linear combination of the forecasts). The parameters of the target peak are identified as a linear regression from the corresponding parameters in forecast sources (markers (4) are at points $(T_L, 120)$, $(T, H)$, $(T_R, 120)$ for target peak). Then each peak within the data sources is shifted to target $T$ (5). After the linear combination of shifted forecasts (6) the ensemble peak is multiplied to fit the target peak height (7). The proposed approach allows to enhance the peak (and thus – floods) forecasting. Fig. 9 (right) shows the percentage of improvement of the main quality metrics: mean average error (MAE), dynamic time wrapping distance to observations (DTW), mean average error for part of the forecast higher than the threshold value (WMAE), standard deviation for forecast error (STDEV), standard deviation for error in peak parameters $H$ and $T$ error (HSTDEV and TSTDEV).

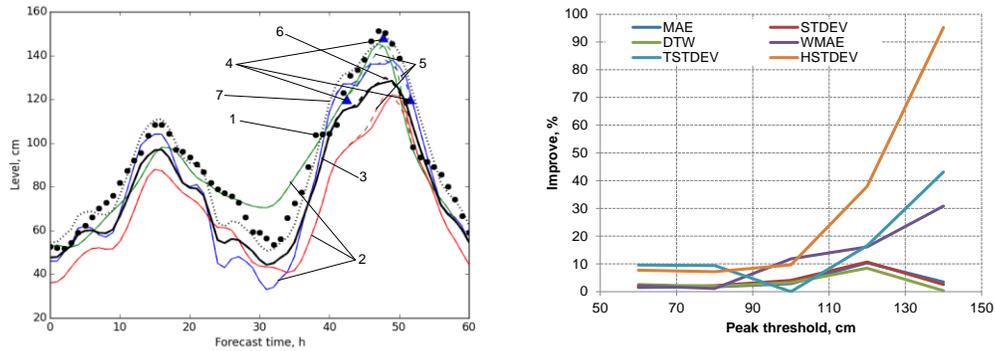

**Figure 9.** Peak parameterization for ensemble improvement:
improvement procedure (left); quality metrics (right)

*Detection of forecast anomalies.* Selection ensemble members (or, in other words, selection of one of the combinatorial ensemble) can be useful to exclude data sources that provide imprecise data. Detection of such data sources can be performed by comparison of available forecasts and excluding those that lie far from others. E.g. fig. 10 shows a case, where detection of outliers in distances between two pairs of three data sources enables identify points, where the particular data source and the ensembles built with it fail and can be excluded.

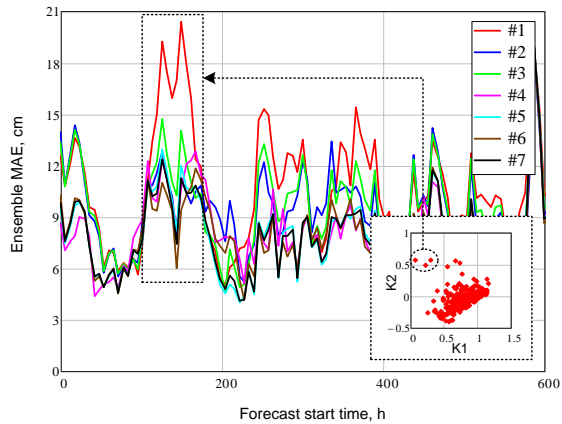

**Figure 10.** Detection of anomalies in forecasts

## 5.4 Cloud computing solution for ensemble-based simulation

In most cases, ensemble-based simulation requires a significant amount of computational resources, which are to be prepared and managed during the execution of appropriate software. To cover these issues, as well as the problems concerned with coupling of models and data sources, we introduce a cloud solution intended to automate and simplify the carrying out of calculations. Generally, the cloud computing infrastructure enables dynamic management of software and hardware resources with a flexible way to scale interoperate and control the services within the composite application, which can be considered as an implementation of simulation ensemble. A composite application contains calls to various cloud services that either provide access to input datasets (observations, external model results, etc.) or to applications (models and other auxiliary software) that are deployed on computing resources managed by the system. The later, which are denoted as internal services, allow to launch the wrapped software with specified parameters, hence giving an opportunity to study their impact on the calculation outcome. This approach is implemented with the use of the CLAVIRE platform [43], which covers most of the issues associated with the heterogeneity of the internal models' system requirements, and the potential complexity of the concomitant calculations. Each of the domain-specific services (that wraps either a piece of applied software or a data source), is provided with a set of basic workflows containing sequences of calls to auxiliary software that allows to provide a unified interface, declared in a high-level domain-oriented description (see fig. 11).

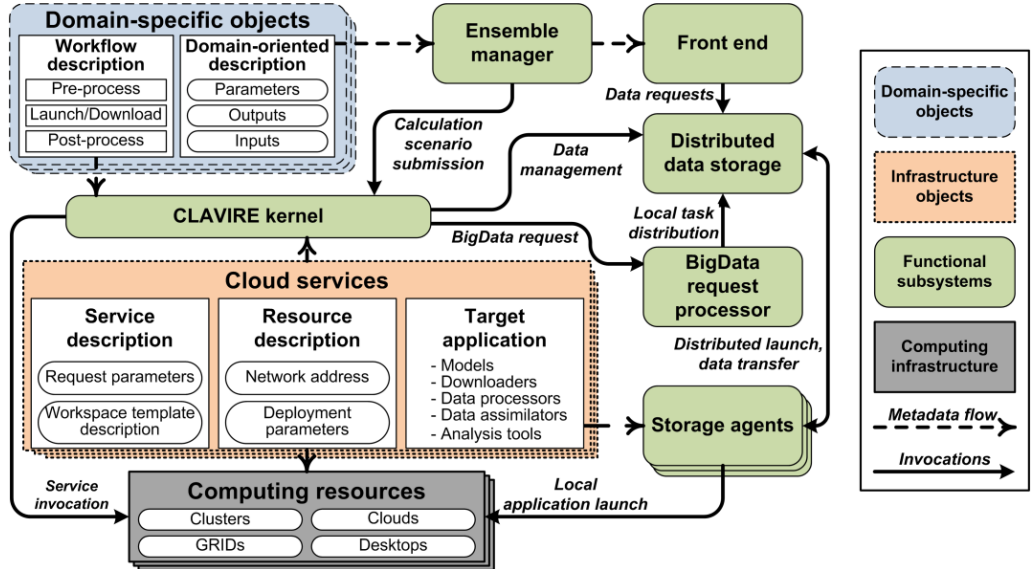

**Figure 11.** Architecture of cloud computing system for ensemble-based simulation

These descriptions formally specify inputs, outputs and parameters of the application in domain terms, which allows the system to identify how the input of one model can be combined with the output of another to construct a data-flow graph. Paths of the graph define pipelines, which represent elements of the ensemble. This procedure is performed by the ensemble manager subsystem, which then translates each pipeline into a script and submits it to the CLAVIRE kernel. Using the service and resource descriptions, the kernel performs scheduling and following parallel execution of the target applications. Depending on the underlying software and corresponding application task, the workflow script may contain BigData requests that are forwarded to the BigData request processor [44]. Further, the tasks are distributed among the storage agents, which perform the direct local launch of the applications. Thus, the cloud computing solution, which is discussed in more details in [45], allows performing an automatic intelligent coupling of models and data sources to take into account all elements of a particular ensemble, especially under the condition when a collection of integrated applied software is regularly expanded.

# 6 Discussion

The proposed general approach can be implemented in various ways. Also, there many ways it can be extended or detailed. One of the most important procedures mentioned in the context of the proposed approach is data assimilation [46]. Besides the basic incorporation of the observations into the simulation process using corresponding capabilities of the model, it can be considered in a more general way as a parameterization of any procedure within the simulation management process (ensemble building, data assessment, classes' identification, ensemble aggregation/selection, etc.).

Considering the evolution of the ensemble over the time the assimilation shape and control its process by comparison to the coming data. This becomes more complicated in case of ensemble-based forecasting as the ensembles on previous time steps often aren't covered by observations. As a result, the maximum quality information is available only for forecasts started as early in past as long is forecasted time.

Next important issue is the selection of the right metrics and quality measures to assess the available data (members of an ensemble or ensembles as a whole) using observations. The complexity of this issue was discussed in Section 4.3. Considering all the variability of existing measures, there is no common and systematic way to select and apply the proper quality measure. Moreover, the right selection of metrics and quality measure require the involvement of domain knowledge (explicitly or implicitly). On the other hand, this knowledge dependency gives us hope that the generalized way of quality metrics selection at least in the area of ensemble-based simulation is possible and even supportable with automatic procedures.

Working with domain knowledge within the proposed approach includes not only explicit expression of the knowledge within the simulation environment, but also the implementation of algorithms that are intended to discover the knowledge within available data (which often related to the involvement of unsupervised machine learning). Nevertheless, this implicit knowledge needs to be controlled using explicit knowledge to avoid overfitting and underfitting of algorithms as well as the possible discovery of well-known facts from the problem domains.

Moreover, the working with domain knowledge become significantly important as the complexity of the simulated system induce the complexity of the simulation system. The simulation system contains diverse data sources, software and hardware resources, which need to be managed within a unified process of simulation. This may require a) application of the proposed principles not only to the simulated system, but also to the simulation system; b) involvement of the knowledge-based technologies to support simulation; c) usage of the specific tools to describe the knowledge and to represent it within the user interface.

# 7   Conclusion and future works

The described approach is developed to extend the conceptual framework for ensemble-based simulation [5] with the use of classification methods as a core for ensemble management and aggregation. It enables a broad range of implementation variants and still keep the core idea of class set identification and selection procedures as a basic loop of the ensemble evolution. The demonstrated forecast-based interpretation and series of its applications shows the capability of the approach to solve the task of quality improvement for ensemble-based simulation of the complex systems.

Still the development of the framework extension is an ongoing project, and there are many directions that are considered as a future works. Some of them are as follows. Application of symbolic regression shows quite interesting results, but it needs to be developed further with hiring domain knowledge. Application of artificial neural networks promises to be a powerful solution for automatically discovered selection functions. Deeper investigation on machine learning algorithms for data analysis and discovering knowledge can provide new ways for ensemble management, classification, and selection within ensemble-based simulations of the complex systems. Development of generalized approach to appropriate metrics and quality measures selection is also considered as an issue for further research.

*Acknowledgements:* This paper is supported by Russian Scientific Foundation, grant #14-11-00823. The research is performed in Advanced Computing Lab (ITMO University), which is created in the frame of 220 Decree of Russian Government, contract #11.G34.31.0019.